# TÌM KIẾM NGỮ NGHĨA SỬ DỤNG KÍCH HOẠT LAN TRUYỀN TRÊN ONTOLOGY

NGÔ MINH VƯƠNG*

**TÓM TẮT**

*Các hệ thống truy hồi tài liệu dạng văn bản hiện nay gặp nhiều thách thức trong việc khám phá ngữ nghĩa của truy vấn và tài liệu. Mỗi truy vấn hàm ý các thông tin tuy không xuất hiện trong truy vấn nhưng các tài liệu nói về các thông tin này cũng nằm trong mong đợi của người đặt truy vấn. Nhược điểm của các phương pháp kích hoạt lan truyền trước đây là có thể có nhiều khái niệm không liên quan được thêm vào truy vấn. Phương pháp mới mà công trình đề xuất là chỉ kích hoạt và thêm vào truy vấn các thực thể có tên có quan hệ với các thực thể xuất hiện trong truy vấn theo các quan hệ tường minh trong truy vấn đó.*

***Từ khóa:*** Ontology, kích hoạt lan truyền, truy hồi tài liệu, mở rộng truy vấn, tìm kiếm ngữ nghĩa.

**ABSTRACT**

***Semantic Search using Spreading Activation based on Ontology***

*Currently, the text document retrieval systems have many challenges in exploring the semantics of queries and documents. Each query implies information which does not appear in the query but the documents related with the information are also expected by user. The disadvantage of the previous spreading activation algorithms could be many irrelevant concepts added to the query. In this paper, a proposed novel algorithm is only activate and add to the query named entities which are related with original entities in the query and explicit relations in the query.*

## 1. Giới thiệu

Ngày nay, nhiều thông tin hữu ích được lưu trữ trên world wide web (www) và, theo bản báo cáo tháng 12/2010 của ITU[1], có hơn 2 tỉ người sử dụng Internet với tần suất thường xuyên. Vì thế, nhu cầu khai thác và sử dụng thông tin trên www một cách hiệu quả là rất lớn. Trong khi đó, một truy vấn thường ngắn gọn, đôi khi diễn đạt không chính xác về một nhu cầu thông tin cơ bản [69]. Để truy vấn có nội dung được diễn đạt rõ nghĩa hơn, phương pháp mở rộng truy vấn được sử dụng rộng rãi trong cộng đồng truy hồi thông tin. Mở rộng truy vấn thông thường là làm tăng độ đầy đủ [66], [26] và đôi khi độ chính xác cũng được cải thiện. [46]

_______________________
* TS, Trường Đại học Tôn Đức Thắng

Theo [79], có hai nhóm phương pháp mở rộng truy vấn. Nhóm một là các phương pháp dựa trên các quan hệ phi cấu trúc. Chúng được rút trích từ sự phân tích tập tài liệu hoặc nhật kí của người dùng. Nhóm hai là các phương pháp dựa trên các quan hệ có cấu trúc. Chúng được rút trích từ các nguồn được biên tập cẩn thận bởi con người. Nhóm một bao gồm các phương pháp như: (1) phản hồi sự liên quan (relevance feedback) [61], [36]; (2) phản hồi sự liên quan giả (pseudo-relevance feedback) [52], [48]; (3) sử dụng nhật kí người dùng [77], [14]; và (4) sử dụng sự đồng xuất hiện [53], [33].

Nhóm hai bao gồm các phương pháp như: (1) khai thác các quan hệ đồng nghĩa, nghĩa cha, nghĩa con trong WordNet [75], [38]; (2) khai thác các quan hệ bí danh, lớp cha, lớp con trong ontology về thực thể có tên [51], [23]; (3) khai thác tất cả các quan hệ trong ontology [59], [40]; và (4) khai thác quan hệ được thể hiện trong truy vấn [31].

Mở rộng truy vấn là thêm vào truy vấn các khái niệm tiềm ẩn không xuất hiện ở truy vấn nhưng góp phần thể hiện rõ nghĩa của truy vấn, hay nói cách khác là làm rõ mong muốn của người dùng. Với nhận định bằng trực giác, việc thêm các khái niệm phù hợp với mục đích của người dùng sẽ tăng độ đầy đủ và độ chính xác của tìm kiếm. Ngược lại, việc thêm các khái niệm không phù hợp sẽ làm giảm hiệu quả truy hồi của hệ thống. Ví dụ với các truy vấn như: (1) tìm kiếm các tài liệu về "*cities that are tourist destinations of Thailand*"; (2) tìm kiếm các tài liệu về "*tsunami in Southeast Asia*"; và (3) tìm kiếm các tài liệu về "*settlements are built in east Jerusalem*"; Ở truy vấn thứ nhất, *Chiang Mai* và *Phuket* nên được thêm vào truy vấn, bởi vì chúng thuộc về lớp *City* và là *tourist destinations of Thailand*. Ở truy vấn thứ hai, các quốc gia có quan hệ "*is part of*" với *Southeast Asia* nên được thêm vào truy vấn này, ví dụ như *Indonesia* hoặc *Philippine*. Tuy nhiên, các quốc gia được thêm vào phải thực sự bị tấn công bởi sóng thần ít nhất một lần. Do đó, *Laos* không được thêm vào truy vấn vì quốc gia nay chưa từng bị sóng thần. Ở truy vấn thứ ba, nếu có sự kiện thể hiện các vị trí mà ở đó các khu tái định cư được xây dựng (*settlements are built in*) và chúng ở phía đông của thành phống *Jerusalem* (*east of Jerusalem*) như *Gilo*, thì vùng đất này nên được thêm vào truy vấn.

Có hai kiểu tìm kiếm trong truy hồi thông tin, đó là truy hồi tài liệu (Document Retrieval) và hỏi–đáp (Question-and-Answering). Chúng được đề cập lần lượt như tìm kiếm rộng khắp (Navigational Search) và tìm kiếm chuyên sâu (Research Search) ở [35]. Hệ thống hỏi-đáp là hệ thống khi được người dùng cung cấp một cụm từ hoặc một câu thì nó trả về các đối tượng, là sự trả lời cho các câu hỏi của người dùng, không phải là các tài liệu. Trong thực tế, các trả lời từ hệ thống hỏi–đáp có thể được sử dụng để tìm kiếm tốt hơn các tài liệu cho các câu hỏi này [29]. Công việc của chúng tôi là truy hồi tài liệu, nghĩa là người dùng cung cấp cho hệ thống một cụm từ hoặc một câu để tìm kiếm các tài liệu mong muốn. Các tài liệu trả về không cần chứa các thuật ngữ ở truy vấn và có thể được xếp hạng bởi sự liên quan của chúng với truy vấn.

Cho đến nay, không có mô hình truy hồi tài liệu nào mở rộng truy vấn trong tài liệu tham khảo có sử dụng quan hệ được thể hiện trong truy vấn để mở rộng truy vấn một cách tổng quát như chúng tôi. Trong công trình này, chúng tôi đề xuất một mô hình không gian vectơ dựa trên ontology. Nó khai thác quan hệ bí danh, lớp cha, lớp con trên các ontology về khái niệm, và kích hoạt lan truyền trên ontology sự kiện theo các quan hệ được thể hiện trong truy vấn. Phần còn lại của công trình được tổ chức như sau. Phần 2 trình bày cơ sở kiến thức và công trình liên quan. Phần 3 giới thiệu về giải thuật kích hoạt lan truyền. Phần 4 mô tả kiến trúc và phương pháp mở rộng truy vấn bằng giải thuật kích hoạt lan truyền có ràng buộc quan hệ. Phần 5 trình bày sự đánh giá và thảo luận trên các kết quả thí nghiệm. Phần cuối cùng là kết luận.

## 2. Cơ sở kiến thức

### 2.1. *Ontology*

- ***Khái niệm***

Ontology bắt nguồn từ triết học, được dẫn xuất từ tiếng Hi Lạp là "onto" và "logia". Trong ngữ cảnh triết học, ontology thuộc một nhánh của siêu hình học, được sử dụng để nghiên cứu về bản chất của sự tồn tại, xác định các sự vật nào thực sự tồn tại và cách thức mô tả chúng (2). Trong những năm gần đây, ontology được sử dụng nhiều trong khoa học máy tính và được định nghĩa khác với nghĩa ban đầu. Theo đó ontology là sự mô hình hóa và đặc tả các các khái niệm một cách hình thức, rõ ràng và chia sẻ được [34], [28]. Thêm vào đó, theo [24], ontology cần có thêm tính thống nhất, tính mở rộng và tính suy luận.

Ontology được sử dụng trong các lĩnh vực như biểu diễn tri thức, xử lí ngôn ngữ tự nhiên, rút trích thông tin, cở sở dữ liệu, quản lí tri thức, các cơ sở dữ liệu trên mạng, thư viện điện tử, hệ thống thông tin địa lí. Các ontology đó có thể chia thính ba nhóm. Nhóm thứ nhất là các ontology được xây dựng thủ công bởi một nhóm các chuyên gia, như WordNet hoặc KIM [42]. Nội dung thông tin trong các ontology này được đầu tư bài bản và kiểm duyệt kĩ lưỡng, do đó có độ tin cậy cao. Tuy nhiên kích thước, mức độ bao phủ và tần suất cập nhật thông tin của chúng bị giới hạn. Nhóm thứ hai là các ontology được xây dựng tự động, ví dụ như YAGO [70], DBpedia (4). Các ontology được phát triển tự động, không tốn nhiều công sức, tuy nhiên chúng có độ tin cậy không bằng các ontology được tạo bởi các chuyên gia. Nhóm thứ ba là các ontology nội dung mở. Ở nhóm này, mọi người đều có thể tham gia đóng góp nội dung thông tin. Điển hình là Wikipedia, từ điển được sử dụng rộng rãi nhất hiện nay trên Internet.

Hệ thống KIM[1] (Knowledge and Information Management) có chứa KIM ontology và cơ sở tri thức (knowledge base – KB)([58]). KIM Ontology định nghĩa các lớp thực thể là các lớp như Person, Organization, Company, Location, và định nghĩa cây phân cấp, các thuộc tính của các lớp thực thể và các quan hệ giữa các lớp thực thể. Ontology của KIM chứa khoảng 300 lớp thực thể, và 100 thuộc tính và kiểu quan hệ. Cơ sở tri thức của KIM chứa đựng thông tin về các thực thể cụ thể thuộc về các lớp thực thể đã được định nghĩa bởi KIM ontology. Hiện nay, KIM có khoảng 77.500 thực thể có tên với hơn 110.000 bí danh được lưu trữ trong cơ sở tri thức của nó.

YAGO (Yet Another Great Ontology) [70], [71] chứa khoảng 1,95 triệu thực thể, 93 kiểu quan hệ và 19 triệu sự kiện mô tả các quan hệ giữa các thực thể. Các sự kiện này được rút trích từ Wikipedia và kết hợp với WordNet bằng cách sử dụng các luật và heuristic. Các sự kiện mới được kiểm tra và thêm vào cở sở tri thức bởi bộ phận kiểm tra của YAGO. Độ chính xác của các sự kiện này là khoảng 95%. Tất cả các đối tượng (ví dụ như thành phố, con người, URLs) được thể hiện như là các thực thể và chúng được liên kết với nhau thông qua các quan hệ.

Wikipedia[2] được xây dựng vào năm 2001 với mục đích tạo ra các bách khoa toàn thư gồm nhiều ngôn ngữ. Ngày nay, nó là một bách khoa toàn thư lớn nhất và được sử dụng nhiều nhất. Wikipedia đã trở thành một hiện tượng trong khoa học máy tính cũng như trong công chúng, với hơn 400 triệu lượt truy cập hàng tháng. Chỉ riêng ở ngôn ngữ tiếng Anh, tính đến này 02 tháng 08 năm 2011, wikipedia có xấp xỉ 3,7 triệu đề mục với hơn 24 triệu trang[3]. Tuy được xây dựng từ các tình nguyện viên, nội dung trên Wikipedia vẫn có chất lượng và độ tin cậy cao như các bài viết tương tự trên Từ điển Bách khoa toàn thư Britannica[4] [32]. Wikipedia có thể được xem như là một Từ điển Bách khoa toàn thư, một từ điển hoặc một ontology. [50]

## 2.2. *Phương pháp kích hoạt lan truyền*

Trong khoa học máy tính, phương pháp SA (Spreading Activation, kích hoạt lan truyền) [21] được sử dụng lần đầu tiên trong lĩnh vực trí tuệ nhân tạo. Gần đây, phương pháp này đã được sử dụng rộng rãi trong truy hồi tài liệu. Phương pháp SA sử dụng một ontology và một số kĩ thuật áp dụng trên ontology này để tìm các khái niệm có liên quan đến truy vấn của người dùng. Ý tưởng cơ bản ẩn bên dưới phương pháp SA là sự khai thác các mối quan hệ giữa các khái niệm trong ontology. Trong đó, các quan hệ thường được đánh nhãn, đánh trọng số, và có thể có hướng.

Trước tiên, phương pháp SA tạo ra một tập khái niệm khởi động từ truy vấn và gán trọng số cho các khái niệm này. Tiếp theo, từ các khái niệm ban đầu, một tập các khái niệm liên quan được tìm kiếm bằng cách lan truyền theo các quan hệ trong ontology. Sau khi các khái niệm gần với các khái niệm ban đầu nhất được kích hoạt, sự kích hoạt sẽ truyền tới các khái niệm tiếp theo trong ontology thông qua các quan hệ trong đó. Sự lan truyền sẽ dừng lại khi một trong các điều kiện kết thúc xảy ra. Các khái niệm được kích hoạt sẽ được gán trọng số và thêm vào truy vấn ban đầu.

Phương pháp SA tự do là phương pháp kích hoạt lan truyền cơ bản nhất. Phương pháp này kích hoạt tất cả các khái niệm có liên quan đến khái niệm ban đầu của truy vấn, thông qua các quan hệ trực tiếp hoặc gián tiếp với khái niệm ban đầu đó trong ontology được sử dụng. Vì sự lan truyền sâu và rộng như thế trên ontology, nhược điểm của phương pháp SA tự do là các khái niệm được kích hoạt phần lớn không liên quan đến nội dung của truy vấn. Điều này làm cho phần lớn các tài liệu trả về bởi phương pháp SA tự do không phù hợp với truy vấn. [8]

Nhược điểm của phương pháp SA tự do có thể được khắc phục một phần bằng cách sử dụng một số luật giới hạn sự lan truyền. Trong phương pháp SA có ràng buộc (Constrained Spreading Activation - CSA), sự lan truyền được giới hạn bởi một số ràng buộc như ràng buộc theo khoảng cách (distance), theo số lượng khái niệm được kích hoạt (fan-out), theo đường dẫn (path), và theo sự kích hoạt (activation). Hình 1 minh họa một phần của một ontology về sự kiện, kết hợp YAGO với Wikipedia, có chứa khái niệm *Thailand*. Với truy vấn tìm kiếm các tài liệu về "*cities that are tourist destinations of Thailand*", căn cứ vào nội dung của truy vấn và các sự kiện được mô tả ở Hình 1, chỉ có hai khái niệm là *Phuket* và *Chiang Mai* cần được kích hoạt và thêm vào truy vấn. Trong khi đó, với phương pháp SA tự do, từ khái niệm *Thailand* ban đầu, mười khái niệm là *Phuket*, *Thaksin Shinawatra*, *Thai Rak Thai*, *Southeast Asia, Vietnam*, *Hanoi*, *Chiang Mai*, *1296*, *Wat Chiang Man*, và *Phang Nga Bay* sẽ được kích hoạt và thêm vào truy vấn; tức là có tám khái niệm không phù hợp được thêm vào truy vấn.

Trong khi đó, với phương pháp SA có ràng buộc về khoảng cách là 1, tức chỉ tính các khái niệm có quan hệ trực tiếp với khái niệm ban đầu, thì có năm khái niệm là *Phuket*, *Thaksin Shinawatra*, *Southeast Asia*, *Chiang Mai*, và *Phang Nga Bay* được kích hoạt và thêm vào truy vấn. Trong đó, *Thaksin Shinawatra* và *Southeast Asia* là không phù hợp vì không phải là điểm đến du lịch của Thái Lan, và *Phang Nga Bay* cũng không phù hợp vì là một điểm đến du lịch nhưng không phải là một thành phố của Thái Lan.

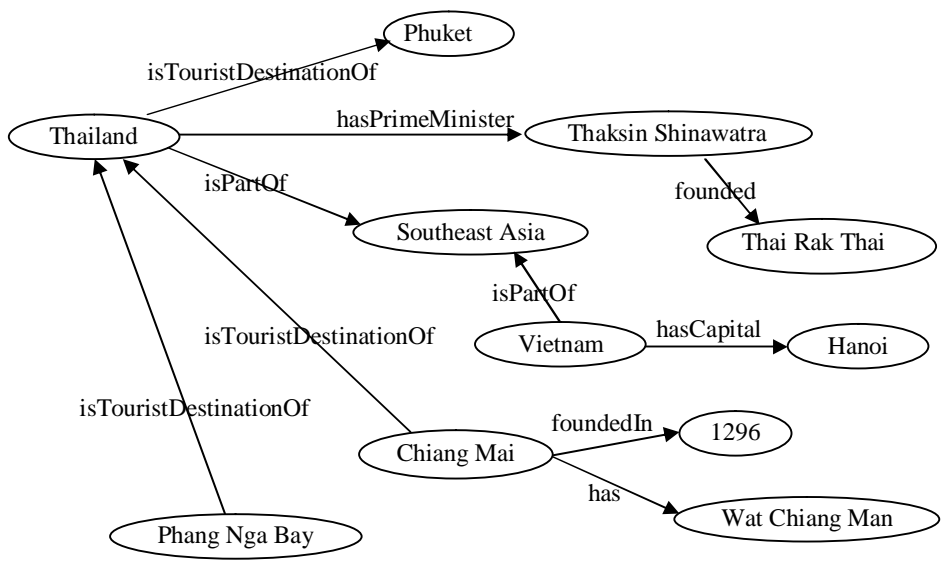

***Hình 1***. *Ví dụ về các khái niệm có liên quan với khái niệm Thailand trong một ontology về sự kiện*

**3.   Các công trình liên quan**

Tìm kiếm ngữ nghĩa, một ứng dụng của Web ngữ nghĩa trong lĩnh vực truy hồi thông tin, đã thể hiện năng lực vượt trội trong việc cải tiến hiệu quả truy hồi. So với các động cơ tìm kiếm truyền thống là tập trung vào đếm tần số xuất hiện của từ, các động cơ tìm kiếm ngữ nghĩa cố gắng hiểu nghĩa tiềm ẩn bên trong của các yêu cầu người dùng và của các thông tin phản hồi. Qua khảo sát và dựa vào sự phân loại ở các công trình trước đó như [49], [25], và [27], chúng tôi nhận thấy tìm kiếm ngữ nghĩa được ứng dụng phần lớn trong các lĩnh vực sau:

1.   Tìm kiếm dựa trên giao diện người dùng theo ngữ nghĩa (Semantic user interface based Search): đây là hệ thống tìm kiếm các thông tin theo truy vấn ban đầu, người dùng dựa vào các thông tin này và chọn thông tin bổ sung cho truy vấn ban đầu của mình. Hệ thống dựa vào đó sẽ tìm kiếm hoặc sắp xếp lại các thông tin trả về cho người dùng. Như các công trình: [16], [1], [22] và [74].

2.   Tìm kiếm hỏi đáp (Question Answering Search): là hệ thống tìm kiếm các trả lời tương ứng cho một câu hỏi hơn là các tài liệu chứa câu trả lời [76]. Có các công trình: [73], [17], [62] và [15].

3.   Xếp hạng thực thể (Entity Ranking): là hệ thống tìm kiếm danh sách các thực thể thuộc một kiểu chính xác và có thể có các tài liệu liên quan với truy vấn thể hiện các thực thể này (5). Ở loại hình này, người dùng muốn tìm kiếm các thực thể được thể hiện trực tiếp bằng một danh sách các thực thể được xếp hạng hơn là một danh sách các trang web không chỉ liên quan với truy vấn mà còn chứa thông tin về các thực thể này. Có các công trình [9], [10], [39] và [78].

4.   Truy hồi thông tin đa ngôn ngữ (Cross-Language Information Retrieval): là hệ thống truy hồi thông tin được viết dưới dạng một ngôn ngữ khác với ngôn ngữ được thể hiện ở truy vấn [64]. Một số công trình như [20], [68], [60], [80] và [18].

5.   Truy hồi tài liệu ngôn ngữ có cấu trúc (Structured Language Document Search): là hệ thống sử dụng các ngôn ngữ có cấu trúc để thể hiện truy vấn và tài liệu. Ví dụ như sử dụng ngôn ngữ RDF: [41], [56], [30], [43] và [37]. Hoặc sử dụng ngôn ngữ XML: [57], [44], [47], [67] và [72].

6.   Truy hồi tài liệu ngôn ngữ tự nhiên (Natural Language Document Search): là hệ thống sử dụng ngôn ngữ tự nhiên để thể hiện truy vấn, và các tài liệu truy hồi được viết bởi các ngôn ngữ tự nhiên. Trong quá trình tìm kiếm, các truy vấn và tài liệu có thể được chú giải ngữ nghĩa, và các tài liệu trả về sẽ được xếp hạng theo độ liên quan với truy vấn. Một số công trình là: [51], [54] và [12]. Mô hình của chúng tôi trình bày ở công trình này là truy hồi tài liệu ngôn ngữ tự nhiên bằng phương pháp kích hoạt lan truyền có ràng buộc theo truy vấn.

Các hệ thống sử dụng sử dụng giải thuật kích hoạt lan truyền (Spreading Activation, SA) để mở rộng truy vấn như [59], [3], [65], [38], [40] và [45]. Tuy nhiên, các hệ thống này không sử dụng các quan hệ trong một truy vấn cho trước để ràng buộc sự lan truyền. Trong khi đó, phương pháp kích hoạt lan truyền ràng buộc quan hệ (relation and distance constrained spreading activation, R&D-CSA) của chúng tôi chỉ kích hoạt các khái niệm có liên quan đến các khái niệm và các quan hệ trong truy vấn.

Trong [59], các tác giả đề xuất một giải thuật kích hoạt lan truyền lai (hybrid), nó kết hợp giải thuật SA với truy hồi thông tin dựa trên ontology. Giải thuật này cho phép người dùng thể hiện truy vấn của họ dưới dạng các từ khóa và tìm các khái niệm trong ontology có các từ khóa này xuất hiện trong sự mô tả của các khái niệm đó. Các khái niệm tìm được sẽ được xem như các khái niệm ban đầu. Các liên kết giữa các khái niệm này với các khái niệm khác trong ontology được gán trọng số và độ lớn của trọng số phụ thuộc vào kiểu của mối liên kết. Sau đó, giải thuật SA được sử dụng để tìm các khái niệm liên quan với các khái niệm được khởi tạo trong ontology. Trong [3], hệ thống sử dụng một mạng SA hai cấp độ để kích hoạt một cách khẳng định hoặc phủ định các khái niệm phù hợp hoặc không phù hợp với các khái niệm ở truy vấn dựa trên các kết quả tìm kiếm theo từ khóa. Hệ thống này cũng sử dụng tập đồng nghĩa của các khái niệm của truy vấn ban đầu để kích hoạt lan truyền, và sử dụng phương pháp máy học sử dụng vectơ hỗ trợ (Support Vector Machine) để huấn luyện và phân loại dữ liệu ở các tài liệu trả về. Trong [65], hệ thống tìm câu trả lời cho câu hỏi và thêm vào câu hỏi này. Sau đó, hệ thống sử dụng giải thuật SA để tìm các khái niệm liên quan đến truy vấn được mở rộng này.

Công trình [38], mở rộng truy vấn bằng cách sử dụng giải thuật SA trên tất cả các quan hệ ở WordNet và chỉ chọn các từ được kích hoạt có bổ sung nghĩa cho nội dung của truy vấn thông qua một số luật. Trong [40], các tác giả không yêu cầu người dùng mô tả các khái niệm trong truy vấn của họ. Hệ thống ánh xạ truy vấn ban đầu thành tập từ khóa và tìm kiếm các tài liệu liên quan với tập từ khóa này. Sau đó, các tài liệu này sẽ được chú giải với các thông tin của ontology và các khái niệm khởi tạo được rút trích từ chúng. Một giải thuật SA được sử dụng để tìm các khái niệm liên quan với các khái niệm được khởi tạo trong ontology. Cuối cùng, các khái niệm được kích hoạt này sẽ được sử dụng để xếp hạng lại các tài liệu để chúng phù hợp hơn với tập từ khóa ban đầu. Trong [45], hệ thống thiết lập một mạng kết hợp với các nút là các trang web và các liên kết giữa các nút là các liên kết giữa các trang web tương ứng. Các nút khởi tạo của giải thuật SA là các trang web có liên quan mạnh với truy vấn cho trước. Tiếp theo, các nút khác (các trang web) sẽ được kích hoạt và trả về cho người dùng.

Một số hệ thống cải thiện hiệu quả truy hồi tài liệu bằng cách mở rộng truy vấn với sự tham gia của người dùng như [63], [6], [14], [52], và [1]. Trong [63], từ các tài liệu liên quan với truy vấn ban đầu, hệ thống đưa ra một cây phân cấp các khái niệm để người dùng chọn và đưa vào truy vấn. Trong 6, các tác giả đề xuất một phương pháp chọn các thuật ngữ thêm vào truy vấn nhưng độc lập với truy vấn bằng cách dựa trên các tài liệu được mô tả bởi người dùng phản ảnh thông tin họ cần nhưng các tài liệu này không được truy hồi bởi truy vấn này. Trong [14], hệ thống khai thác nhật kí truy vấn của người dùng để liệt kê các ứng viên đồng nghĩa phù hợp với truy vấn ban đầu. Trong đó, nhật kí truy vấn của người dùng là các truy vấn đăng nhập, các kết quả tìm kiếm được xem và các URL được nhấp chuột. Từ danh sách ứng viên này, người dùng sẽ chọn ứng viên phù hợp trong ngữ cảnh của một cơ sở tri thức. Trong [52], hệ thống mở rộng truy vấn bằng cách chọn thông tin trong tất cả tài liệu trả về cho truy vấn ban đầu và thông tin của các tài liệu được người dùng đánh giá để thêm vào truy vấn. Trong [1], hệ thống rút trích các thực thể có tên từ tập tài liệu trả về cho truy vấn ban đầu. Tiếp theo, người dùng sẽ chọn các thực thể có tên phù hợp để thêm vào truy vấn. Bên cạnh đó, [7] cô động nội dung của truy vấn bằng cách loại bỏ các khái niệm thể hiện thông tin không quan trọng trong truy vấn. Trong khi, hệ thống của chúng tôi tiến hành mở rộng truy vấn một cách tự động.

Một số hệ thống khác mở rộng truy vấn bằng cách sử dụng thông tin được lưu trữ trong ontology như [73], [17] và [13]. Trong [73], các tác giả ánh xạ các khái niệm của truy vấn vào trong ontology để tìm các khái niệm liên quan phù hợp. Trong [17], mục tiêu của hệ thống là tìm kiếm các thực thể có tên thuộc các lớp được mô tả kết hợp với từ khóa trong truy vấn. Tuy nhiên, hai công trình này không khảo sát tới các quan hệ trong truy vấn và chúng ứng dụng cho hệ thống hỏi-đáp chứ không phải cho truy hồi tài liệu. Trong [13], hệ thống tìm các thực thể có tên xác định thuộc một lớp thực thể có tên trong truy vấn, sau đó vectơ của truy vấn sẽ được khởi tạo từ các thực thể có tên này. Bước này làm tốn thời gian không cần thiết. Hơn nữa, một cơ sở tri thức thường không đầy đủ, nên các tài liệu phù hợp chứa các thực thể có tên không tồn tại trong cơ sở tri thức sẽ không được trả về. Trong mô hình của chúng tôi, các vectơ truy vấn và tài liệu có chứa lớp thực thể có tên này sẽ được khởi tạo và so khớp ngay. Bên cạnh đó, các truy vấn của công trình trên phải được mô tả ở dạng RDQL.

Ở [55], hệ thống chuyển truy vấn thành cụm danh từ bao gồm đối tượng, thành phần của đối tượng và tính chất của thành phần. Tác giả đề xuất hai phương pháp mở rộng truy vấn. Phương pháp thứ nhất là tìm kiếm các cụm danh từ tương tự với cụm danh từ ban đầu trong ontology về cụm danh từ của tác giả tự xây dựng. Ở phương pháp thứ hai, tác giả có sử dụng thêm kĩ thuật phản hồi liên quan. Giải thuật phản hồi liên quan giả mở rộng truy vấn bằng cách sử dụng các thuật ngữ trong các tài liệu có thứ hạng cao trong lần truy hồi với truy vấn ban đầu. Giải thuật này làm tiêu tốn thời gian do phải truy vấn hai lần, điều này làm giới hạn ứng dụng của nó trong thực tế. Cụ thể là từ các tài liệu liên quan với truy vấn ban đầu, hệ thống này sẽ tìm kiếm các cụm danh từ có mối quan hệ trong ontoloy về cụm danh từ của tác giả với cụm danh từ ban đầu trong truy vấn để thêm vào truy vấn. Cả hai phương pháp đều không sử dụng mối quan hệ trong truy vấn và tác giả chỉ giới hạn ở các truy vấn chuyển được về dạng cụm danh từ gồm đối tượng, tính chất và thành phần.

Công trình [31], các tác giả có sử dụng các quan hệ trong truy vấn để mở rộng nó. Tuy nhiên, công trình này chỉ khai thác các quan hệ không gian (ví dụ: *near*, *inside*, *north of*). Ngược lại, chúng tôi đề xuất các luật tổng quát hơn cho mở rộng truy vấn. Bên cạnh đó, trong [77], hệ thống sử dụng các quan hệ đồng nghĩa hoặc đồng xuất hiện trong nhật kí truy vấn của người dùng để chỉnh sửa hoặc mở rộng truy vấn. Trong [41], các truy vấn phải được viết dưới dạng SPARQL. Các khái niệm và quan hệ phải được mô tả rõ ràng bởi người dùng. Điều này sẽ gây khó khăn cho người sử dụng. Hơn nữa, công trình này dành cho hệ thống hỏi-đáp chứ không dành cho truy hồi tài liệu. Trong [48], hệ thống kết hợp giải thuật phản hồi liên quan giả với kĩ thuật phân tích nội dung cục bộ để mở rộng truy vấn.

**4. Mở rộng truy vấn**

Phương pháp kích hoạt lan truyền để mở rộng truy vấn mà chúng tôi đề xuất trong công trình này là phương pháp ràng buộc theo quan hệ, được gọi là R+CSA. Kiến trúc hệ thống sử dụng R+CSA được trình bày trong Hình 2. Truy vấn ban đầu được mở rộng thông qua mô đun *Phương pháp R+CSA*. Tiếp theo các tài liệu và truy vấn mở rộng sẽ được biểu diễn bởi các không gian vectơ dựa trên từ khóa. Cuối cùng, việc lọc và xếp hạng tài liệu được thực hiện như với mô hình không gian vec tơ truyền thống (Vector Space Model, VSM) thông qua mô đun *VSM dựa trên từ khóa*, trong đó trọng số của các từ khóa được tính theo *tf.idf*.

Hình 3 trình bày năm bước chính của phương pháp R+CSA để xác định thông tin tiềm ẩn liên quan với truy vấn. Chi tiết của phương pháp R+CSA gồm các bước sau:

1. *Nhận diện quan hệ*: nhận diện các cụm từ quan hệ trong truy vấn và ánh xạ chúng thành các quan hệ tương ứng trong ontology được sử dụng.

2. *Nhận diện các khái niệm khởi động*: nhận diện và chú giải các thực thể xuất hiện trong truy vấn.

3. *Thiết lập các bộ quan hệ*: biểu diễn truy vấn ban đầu thành các bộ quan hệ *I-R-C* (hoặc *C-R-I*) cho mỗi quan hệ *R* được xác định ở bước 1, với *I* và *C* lần lượt là một thực thể có tên xác định và một lớp thực thể được nhận diện ở bước 2.

Ví dụ với truy vấn "*Where is the actress, Marion Davies, buried?*", cụm từ quan hệ được xác định bởi hai từ "*where*" và "*buried*" được ánh xạ thành quan hệ *R* là *buriedIn*, *Marion Davies* được nhận diện là thực thể có tên có định danh *I* là *#Marion_Davies* và có lớp là *Woman*, và từ "*where*" được ánh xạ thành lớp *C* là *Location*. Vì vậy bộ quan hệ được thiết lập trong truy vấn này là [*I*: *#Marion_Davies*]-(*R*: *buriedIn*)-[*C*: *Location*].

4. *Kích hoạt lan truyền có ràng buộc theo quan hệ tường minh trong truy vấn*: với mỗi bộ quan hệ *I-R-C*, tìm các thực thể có tên tiềm ẩn $I_a$ có quan hệ *R* với *I* và $I_a$ có lớp là *C* hoặc là lớp con của *C* trong ontology. Ví dụ, trong ontology được sử dụng có quan hệ:

[*I*: *#Marion_Davies*]-(*R*: *buriedIn*)-[$I_a$: *#Hollywood_Cemetery*]

và *#Hollywood_Cemetery* là thực thể có lớp là lớp con của *Location*, nên đó là một thực thể có tên tiềm ẩn cần tìm cho bộ quan hệ ví dụ thiết lập ở bước 3.

5. *Mở rộng truy vấn*: thêm vào truy vấn tên chính của mỗi $I_a$ tiềm ẩn được tìm thấy. Ở ví dụ trên, "*Hollywood Cemetery*" được thêm vào truy vấn.

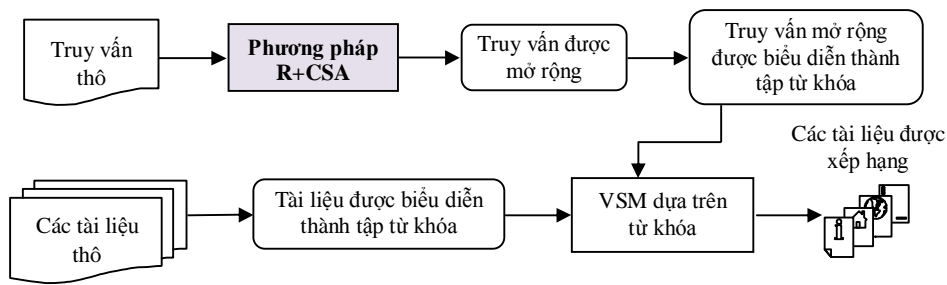

*Hình 2. Kiến trúc hệ thống của mô hình mở rộng truy vấn sử dụng phương pháp R+CSA*

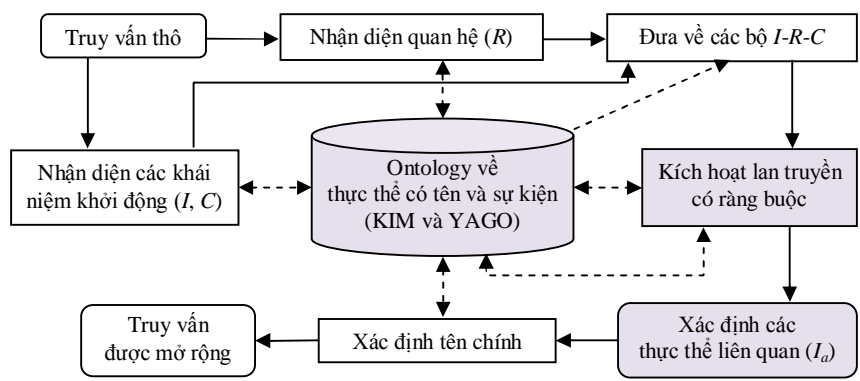

*Hình 3. Các bước của phương pháp R+CSA*

Như vậy, so với phương pháp SA tự do, phương pháp R+CSA có ba ràng buộc. Thứ nhất là ràng buộc về khoảng cách. Tức là, dựa trên ontology về sự kiện được sử dụng, chỉ các thực thể có quan hệ trực tiếp với các thực thể ban đầu xuất hiện trong truy vấn mới được kích hoạt. Thứ hai là ràng buộc về quan hệ; tức là, trên ontology về sự kiện, sự lan truyền chỉ được thực hiện trên các quan hệ xuất hiện tường minh trong truy vấn. Thứ ba là về lớp thực thể; tức là, lớp của mỗi thực thể được kích hoạt phải giống với, hoặc là lớp con của, lớp theo quan hệ tương ứng trong truy vấn.

**5. Đánh giá thực nghiệm**

Để tiến hành thực nghiệm mô hình R+CSA, chúng tôi chọn tập tài liệu L.A. Times và tập truy vấn của QA-Track-99, gồm 124 truy vấn có tài liệu liên quan thuộc tập tài liệu này. Mô hình R+CSA cần sử dụng một ontology có các đặc điểm là: (1) số lượng lớn thực thể có tên; (2) số lượng lớn lớp; (3) hệ thống phân cấp cho các lớp; (4) số lượng lớn quan hệ; (5) các quan hệ hai ngôi có ràng buộc về miền xác định và miền giá trị; và (6) số lượng lớn sự kiện. Tuy nhiên, không có một ontology đơn đủ lớn để bao phủ tất cả các miền và ứng dụng, nói chung, hoặc để đáp ứng yêu cầu về 6 đặc điểm ở trên, nói riêng. Vì vậy, kết hợp nhiều ontology lại với nhau là một giải pháp. [19]

KIM là một ontology tốt về các đặc điểm thứ 1, 2 và 3, tương đối tốt về hai đặc điểm thứ 4 và 5, nhưng không có đặc điểm thứ 6. Trong khi đó, YAGO là một ontology tốt về hai đặc điểm thứ 1 và 6, tương đối tốt về đặc điểm thứ 4, nhưng không tốt về đặc điểm thứ 2 và không có hai đặc điểm thứ 3 và 5. Do đó, để làm thí nghiệm, chúng tôi kết hợp ontology về thực thể có tên của KIM với ontology về sự kiện của YAGO.

Mặc dù vậy, trong 124 truy vấn của QA-Track-99, YAGO chỉ bao phủ được các quan hệ và sự kiện cho 16 truy vấn. Do đó, chúng tôi phải làm giàu thêm YAGO bằng cách: (1) bổ sung thêm 57 quan hệ có trong tập truy vấn nhưng không có trong YAGO, nâng tổng số quan hệ trong YAGO lên thành 150 quan hệ; và (2) tìm trong Wikipedia các sự kiện liên quan đến các thực thể và quan hệ trong tập truy vấn và bổ sung chúng vào YAGO. Mặt khác, chúng tôi cũng phải bổ sung vào KIM ontology các ràng buộc về miền xác định và miền giá trị cho các quan hệ có trong YAGO nhưng không có trong KIM ontology. Với YAGO và KIM ontology được làm giàu như vậy, có tất cả 92 truy vấn mở rộng được theo phương pháp R+CSA, 26 truy vấn không có bộ quan hệ *I-R-C*, và 6 truy vấn không có được sự kiện tương ứng trong YAGO đã làm giàu.

***Bảng 1****. Các độ chính xác và độ F trung bình tại mười một điểm đầy đủ chuẩn của các mô hình Lexical, CSA và R+CSA*

| Độ đo | Mô hình | Độ đầy đủ (%) | | | | | | | | | | |
|---|---|---|---|---|---|---|---|---|---|---|---|---|
| | | 0 | 10 | 20 | 30 | 40 | 50 | 60 | 70 | 80 | 90 | 100 |
| **Độ chính xác** (%) | Lexical | 66,0 | 65,8 | 63,4 | 60,3 | 56,6 | 55,0 | 45,8 | 40,4 | 38,0 | 37,5 | 37,2 |
| | CSA | 68,2 | 67,8 | 66,3 | 63,3 | 60,5 | 59,1 | 50,6 | 47,7 | 46,4 | 44,9 | 44,5 |
| | R+CSA | 78,4 | 77,9 | 75,9 | 73,0 | 69,6 | 68,5 | 61,5 | 57,6 | 55,6 | 54,5 | 53,4 |
| **Độ *F*** (%) | Lexical | 0 | 15,6 | 26,7 | 34,9 | 40,2 | 45,2 | 43,6 | 42,3 | 42,0 | 43,3 | 44,4 |
| | CSA | 0 | 15,3 | 26,7 | 35,1 | 41,4 | 46,9 | 46,5 | 47,4 | 49,1 | 50,1 | 51,6 |
| | R+CSA | 0 | 16,7 | 29,4 | 39,0 | 46,2 | 52,9 | 54,2 | 55,0 | 57,1 | 59,0 | 60,4 |

Về các bước xử lí của phương pháp R+CSA, ở bước 1 để nhận diện và ánh xạ quan hệ, một từ điển ánh xạ các cụm từ quan hệ vào các quan hệ trong ontology được xây dựng trước. Ví dụ, "*actress in*" được ánh xạ thành quan hệ *actedIn* và "*nationality is*" được ánh xạ thành quan hệ *citizenOf* trong YAGO và KIM ontology. Ở bước 2, việc nhận diện các thực thể khởi động trong truy vấn được thực hiện bởi động cơ nhận diện thực thể có tên của KIM có độ chính xác và độ đầy đủ lần lượt vào khoảng 90% và 86%[6]. Việc ánh xạ từ để hỏi đến lớp của thực thể có tên, trong phạm vi bài báo để tiến hành thí nghiệm, được hiện thực thông qua một tập luật đơn giản bao phủ tập dữ liệu kiểm tra. Ở bước 3, phương pháp sinh đồ thị khái niệm trong [11] được áp dụng để kết nối mỗi quan hệ nhận diện được ở bước 1 với các thực thể tương ứng nhận diện được ở bước 2, tạo thành một bộ ba quan hệ. Ở bước 4, với kĩ thuật đánh chỉ mục cho các đối tượng trong một ontology như hiện nay, tìm trong ontology đó một thực thể có quan hệ cho trước với một thực thể cho trước là một tác vụ cơ bản, được thực thi dễ dàng và nhanh.

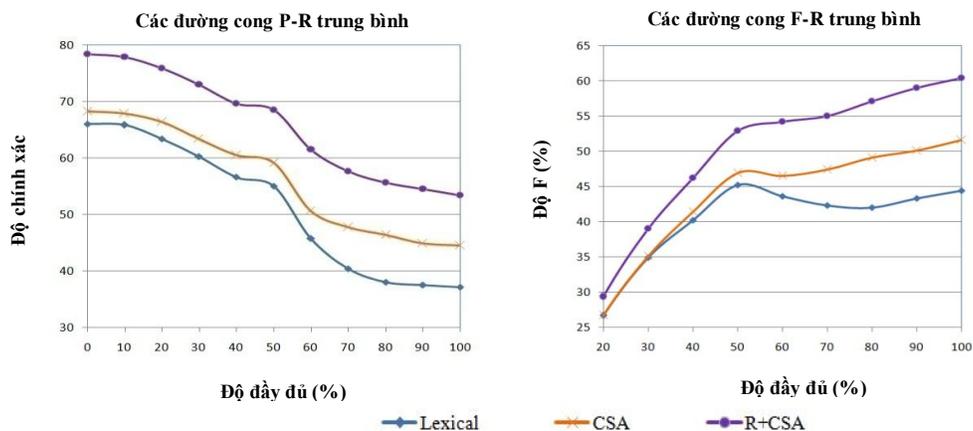

**Hình 4.** *Đường cong trung bình P-R và F-R của các mô hình Lexical, CSA và R+CSA*

Chúng tôi so sánh hiệu quả truy hồi tài liệu giữa mô hình R+CSA đề xuất với hai mô hình sau:

1. Lexical: là mô hình không gian vectơ dựa trên từ khóa truyền thống được hiện thực trong Lucene.

2. CSA: là mô hình sử dụng phương pháp kích hoạt lan truyền có ràng buộc theo khoảng cách. Nó mở rộng truy vấn bằng cách lan truyền trên YAGO (đã làm giàu) theo tất cả các quan hệ trực tiếp với các thực thể ban đầu trong truy vấn. Các truy vấn mở rộng và các tài liệu sau đó cũng được biểu diễn theo mô hình không gian vectơ dựa trên từ khóa.

**Bảng 2.** *Các độ chính xác trung bình nhóm của các mô hình Lexical, CSA và R+CSA*

| Mô hình | R+CSA | Lexical | CSA |
|---|---|---|---|
| **MAP** | **0,6451** | 0,5099 | 0,5474 |
| **Độ cải thiện** | | 26,5% | 17,8% |

Các giá trị trong bảng 1 và các đường cong trong hình 4 trình bày các độ chính xác và độ *F* trung bình của ba mô hình Lexical, CSA và R+CSA tại mỗi cấp độ đầy đủ chuẩn. Chúng cho thấy mô hình R+CSA hiệu quả hơn hai mô hình còn lại ở độ chính xác và độ *F*. Các độ MAP trong Bảng 2 và các trị số *p* hai chiều trong bảng 3 cho thấy việc mở rộng truy vấn một cách hợp lí sẽ làm tăng hiệu quả của truy hồi văn bản. Ở độ MAP, mô hình R+CSA của chúng tôi hiệu quả hơn lần lượt là 26,5% và 17,8% so với hai mô hình Lexical và CSA. Trong khi đó, số các truy vấn mà mô hình R+CSA có độ chính xác trung bình lớn hơn, bằng và nhỏ hơn so với mô hình Lexical lần lượt là 61, 37 và 26; còn so với mô hình CSA lần lượt là 57, 49 và 18.

***Bảng 3****. Trị số p hai chiều của phương pháp kiểm định ngẫu nhiên Fisher giữa mô hình R+CSA với hai mô hình Lexical và CSA*

| **Mô hình *A*** | **Mô hình *B*** | **\|MAP(*A*) – MAP(*B*)\|** | ***N⁻*** | ***N⁺*** | **Trị số *p* hai chiều** |
|---|---|---|---|---|---|
| R+CSA | Lexical | 0,1352 | 1.691 | 1.630 | 0,03321 |
|  | CSA | 0,0977 | 2.207 | 2.268 | 0,04475 |

Dưới đây, chúng tôi trình bày và phân tích một số truy vấn điển hình trong tập QA-Track-99 cho thấy mô hình R+CSA hiệu quả hơn hoặc thua hai mô hình Lexical và CSA, như được trình bày trong Bảng 4. Do mô hình R+CSA là mô hình CSA có ràng buộc thêm mối quan hệ tường minh trong truy vấn, nên các thuật ngữ được thêm vào truy vấn theo mô hình R+CSA cũng được thêm vào truy vấn theo mô hình CSA. Sau đây là các truy vấn và sự phân tích của chúng tôi.

**Truy vấn *a*.** "*What is the capital of Italy?*"

Lexical: *capital* OR *Italy*

CSA: *capital* OR *Italy* OR *Rome* OR *A.S. Roma* OR *A.C. Milan* OR *ACF Fiorentina* OR *Berlusconi* OR *Italian Republic* OR *G8* OR *European Union*

R+CSA: *capital* OR *Italy* OR *Rome*

Bộ quan hệ trong truy vấn này là [*C*: *Capital*]-(*R*: c*apitalOf*)-[*I*: *#Italy*]. Trong ontology về sự kiện có bộ quan hệ tương ứng là [*Iₐ*: *#Rome*]-(*R*: c*apitalOf*)-[*I*: *#Italy*]. Ontology về thực thể có tên xác định [*Iₐ*: *#Rome*] có lớp là [*C*: *Capital*]. Do đó mô hình R+CSA thêm từ khoá "*Rome*" vào truy vấn. Trong khi đó mô hình CSA thêm vào truy vấn các từ khoá biểu diễn bất kỳ thực thể nào có quan hệ với *Italy* trong ontology về sự kiện. Hai mô hình R+CSA và CSA hiệu quả hơn mô hình Lexical vì có một số tài liệu liên quan đến truy vấn có chứa *Rome* mà mô hình Lexical không truy hồi. Mô hình R+CSA hiệu quả hơn mô hình CSA vì mô hình CSA thêm vào truy vấn nhiều từ khoá không phù hợp với nội dung của truy vấn.

**Truy vấn *b*.** "*How many moons does Jupiter have?*"

Lexical: *moon* OR *Jupiter*

CSA: *moon* OR *Jupiter* OR *four* OR *Jupiter Hammerheads* OR *Jupiter Hammon* OR *Jupiter One* OR *Maya Jupiter* OR *Sailor Jupiter* OR *Florida*

R+CSA: *moon* OR *Jupiter* OR *four*

Ở truy vấn này, bộ quan hệ là [*I*: *#Jupiter*]-(*R*: m*oonQuantity*)-[*C*: *Number*]. Trong ontology về sự kiện có bộ quan hệ tương ứng là [*I*: *#Jupiter*]-(*R*: m*oonQuantity*)-[*Iₐ*: *#four*]. Ontology về thực thể có tên xác định [*Iₐ*: *#four*] có lớp là [*C*: *Number*]. Do đó mô hình R+CSA thêm từ khoá "*four*" vào truy vấn. Đây là từ khóa phù hợp với nội dung của truy vấn. Do đó, tương tự như ở truy vấn *a*, ở truy vấn này mô hình R+CSA hiệu quả hơn mô hình CSA và cả hai mô hình này đều hiệu quả hơn mô hình Lexical.

***Bảng 4****. Các độ chính xác trung bình của các mô hình Lexical, CSA và R+CSA trên các truy vấn điển hình*

| Mô hình | Độ chính xác trung bình | | | |
|---|---|---|---|---|
| | *a* | *b* | *c* | *d* |
| Lexical | 0,3929 | 0,1956 | 0,75 | 1 |
| CSA | 0,5071 | 0,4542 | 0,5889 | 0,5 |
| R+CSA | 0,8333 | 0,6496 | 1 | 0,3333 |

**Truy vấn *c*.** "*Where is the actress, Marion Davies, buried?*"

Lexical: *actress* OR *Marion Davies* OR *bury*

CSA: *actress* OR *Marion Davies* OR *bury* OR *Hollywood Cemetery* OR *Blondie of the Follies* OR *Going Hollywood* OR *Janice Meredith* OR *Lights of Old Broadway* OR *Zander the Great* OR *Patricia Lake* OR *Ziegfeld Girls*

R+CSA: *actress* OR *Marion Davies* OR *bury* OR *Hollywood Cemetery*

Ở truy vấn này, mô hình R+CSA khai thác được các bộ quan hệ trong truy vấn và trong ontology về sự kiện lần lượt là [*I*: #*Marion Davies*]-(*R*: *buriedIn*)-[*C*: *Location*], [*I*: #*Marion_Davies*]-(*R*: *buriedIn*)-[$I_a$: #*Hollywood_Cemetery*]. Theo ontology về thực thể có tên, [$I_a$: #*Hollywood_Cemetery*] có lớp là lớp con của [*C*: *Location*]. Do đó, từ "*Hollywood Cemetery*" được thêm vào truy vấn theo mô hình R+CSA. Do đây thật sự là hai từ khoá xuất hiện trong các tài liệu liên quan đến truy vấn nên mô hình R+CSA hiệu quả hơn mô hình Lexical. Trong khi đó mô hình CSA có hiệu quả truy hồi thấp hơn mô hình Lexical, do thêm vào truy vấn quá nhiều từ khoá không phù hợp với nội dung của truy vấn nên có nhiều tài liệu không liên quan đến truy vấn được trả về.

**Truy vấn *d*.** "*What famous communist leader died in Mexico City?*"

Lexical:    *famous* OR *communist* OR *leader* OR *die* OR *Mexico* OR *city*

CSA: *famous* OR *communist* OR *leader* OR *die* OR *Mexico* OR *city* OR *Adolfo Ruiz Cortines* OR *Adolfo de la Huerta* OR *North America* OR *Adolfo Aguilar Zínser* OR *Agustin Carstens* OR *Alejandro Gonzalez Alcocer* OR *Bernardo Gomez Martinez* OR *Alvaro Obregon* OR *Andres Eloy Blanco*

R+CSA: *famous* OR *communist* OR *leader* OR *die* OR *Mexico* OR *city* OR *Adolfo Ruiz Cortines* OR *Adolfo de la Huerta*

Ở truy vấn này, mô hình R+CSA khai thác được bộ quan hệ [*C*: *Leader*]-(*R*: *diedIn*)-[*I*: #*Mexico_City*] trong truy vấn, và các bộ quan hệ [$I_a$: #*Adolfo_Ruiz_Cortines*]-(*R*: *diedIn*)-[*I*: #*Mexico_City*] và [$I_a$: #*Adolfo_de_la_Huerta*]-

149

(*R*: *diedIn*)-[*I*: *#Mexico_City*] trong ontology về sự kiện. Theo ontology về thực thể có tên, [*I$_a$*: *#Adolfo_Ruiz_Cortines*] và [*I$_a$*: *#Adolfo_de_la_Huerta*] có lớp là [*C*: *Leader*]. Do đó, các từ "*Adolfo Ruiz Cortines*" và "*Adolfo de la Huerta*" được thêm vào truy vấn theo mô hình R+CSA. Tuy nhiên, các tài liệu liên quan đến truy vấn chủ yếu chứa thực thể *Leon Trotsky* nhưng ontology sự kiện được sử dụng không có quan hệ [*I$_a$*: *#Leon_Trotsky*]-(*R*: *diedIn*)-[*I*: *#Mexico_City*]. Do đó "*Leon Trotsky*" không được thêm vào truy vấn theo hai mô hình R+CSA và CSA, nên hai mô hình này có hiệu quả thấp hơn mô hình Lexical. Ngoài ra, mô hình R+CSA có hiệu quả thấp hơn mô hình CSA bởi vì có các thực thể xuất hiện trong các tài liệu liên quan đến truy vấn nhưng không được mô hình R+CSA thêm vào truy vấn; đó là các thực thể có quan hệ với các thực thể trong truy vấn nhưng không phải theo các quan hệ tường minh trong truy vấn.

**6. Kết luận**

Công trình đã phân tích các nhược điểm của các phương pháp SA tự do và có ràng buộc trước đây, và đề xuất phương pháp SA có ràng buộc theo quan hệ tường minh trong truy vấn. Cụ thể là, với mỗi truy vấn, mỗi thực thể tiềm ẩn được thêm vào phải liên quan với một thực thể trong truy vấn theo một quan hệ tường minh xuất hiện trong đó, và thuộc lớp của thực thể tương ứng với nó trong truy vấn. Mô hình mở rộng truy vấn theo phương pháp SA đề xuất đã được so sánh về hiệu quả truy hồi tài liệu với mô hình Lexical và mô hình sử dụng phương pháp SA có ràng buộc theo khoảng cách. Kết quả thực nghiệm cho thấy mô hình R+CSA có hiệu quả cao hơn ở độ chính xác, độ đầy đủ, độ *F* và độ MAP.

Trong công trình này, về việc khai thác các thực thể có tên tiềm ẩn, truy vấn chỉ được mở rộng với các thực thể có định danh và thuộc lớp thực thể tham gia trực tiếp vào các quan hệ xuất hiện tường minh trong truy vấn. Một hướng nghiên cứu đáng quan tâm tiếp theo là khai thác các thực thể tiềm ẩn có quan hệ bắc cầu với các thực thể trong truy vấn thông qua các quan hệ tường minh trong đó.

___________________

[1] International Telecommunication Union, là cơ quan chuyên môn của Liên Hợp Quốc về công nghệ thông tin và truyền thông. *http://www.itu.int/net/itunews/issues/2010/10/04.aspx*

[2]  *http://www.ontotext.com/kim/*

[3] *http://www.wikipedia.org/*

[4] *http://en.wikipedia.org/wiki/Wikipedia:About (accessed on 01-Aug-2011)*

[5] *http://www.britannica.com/*

[6] *http://www.ontotext.com/kim/performance.html.*

**TÀI LIỆU THAM KHẢO**